\documentclass[epjCONF]{svjour}
\usepackage{graphics}
\usepackage[varg]{txfonts} 
\usepackage[latin1]{inputenc}
\usepackage{natbib}

\newcommand\araa{{ARA\&A\,}}
\newcommand\apj{{ApJ\,}}
\newcommand\apjl{{ApJ\,}}

\newcommand\aap{{A\&A\,}}
\newcommand\aaps{{A\&AS\,}}
\newcommand\mnras{{MNRAS\,}}
\newcommand\pasp{{PASP\,}}

\session-title{Conference Title, to be filled}
\begin{document}
\title{The SAGA so far: reading the history of the Galaxy with 
asteroseismology}
\author{Luca Casagrande\inst{1}\fnmsep\thanks{\email{luca.casagrande@anu.edu.au}}\fnmsep\thanks{\dag{Stromlo Fellow}} \and Victor Silva Aguirre\inst{2} \and Aldo M. Serenelli\inst{3} \and Dennis Stello\inst{2,4} \and Sofia Feltzing\inst{5} \and Katharine J.\,Schlesinger\inst{1} on behalf of the SAGA\inst{5} team}
\institute{Research School of Astronomy \& Astrophysics, The Australian 
National University, ACT, Australia\and 
Stellar Astrophysics Centre, Department of Physics and Astronomy, Aarhus 
University, Denmark\and 
Institute of Space Sciences, Campus UAB, Fac.\,Ci\'encies, E-08193 Bellaterra, Spain\and
Sydney Institute for Astronomy, School of Physics, University of Sydney, NSW 2006, Australia\and
Lund Observatory, Department of Astronomy and Theoretical Physics, SE-22100 Lund, Sweden\and
http://www.mso.anu.edu.au/saga}
\abstract{
Asteroseismology has the capability of delivering stellar properties 
which would otherwise be inaccessible, such as radii, masses and thus ages of 
stars. When this information is coupled with classical determinations of 
stellar parameters, such as metallicities, effective temperatures and angular 
diameters, powerful new diagnostics for stellar and Galactic studies can 
be obtained. The ongoing Str\"omgren survey for Asteroseismology 
and Galactic Archaeology (SAGA) is pursuing such a goal, by determining 
photometric stellar parameters for stars with seismic oscillations measured by 
the {\it Kepler} satellite. As the survey continues and expands in sample size, 
SAGA will provide an unprecedented opportunity to constrain theories of the 
evolution of the Milky Way disc.
} 
\maketitle
\section{Introduction}
\label{intro}
The study of the formation and evolution of our Galaxy is entering its golden 
age, with a number of spectroscopic and photometric surveys targeting one of its
main (baryonic) components: stars. Among the latter, red giants are the ideal 
targets to decipher the formation history of the Milky Way: on the HR diagram 
they span a vastly different range of gravities and luminosities, thus probing 
a large range of distances. Their ages essentially cover the entire history of 
the Universe, thus making them fossil remnants from different epochs of the 
formation of the Galaxy.
The cold surface temperatures encountered in red giants are the realm of 
interesting atomic and molecular physics shaping their emergent spectra. This 
temperature regime is also dominated by convection, which is the main 
driver of the oscillation modes that we are now able to 
detect in several thousands of stars thanks to space borne asteroseismic 
missions 
such as {\it CoRoT} and {\it Kepler} \citep[e.g.,][for a review]{CM13}.
By measuring oscillation frequencies in stars, asteroseismology allows us to 
measure fundamental physical quantities, masses and radii in particular, which 
otherwise would be inaccessible in single field stars, and which can be used 
to obtain information on stellar distances and ages 
\citep[e.g.,][]{VSA11,VSA12,M13}. 
In particular, global oscillation frequencies not only are the easiest ones to 
detect and analyze, but are also able to provide the aforementioned 
parameters for a large number of stars with an accuracy that is generally much 
better than achievable by isochrone fitting in the traditional sense 
\citep[see e.g.,][]{VSA13,L14a,L14b}.

Asteroseismology thus provides a powerful and new complementary tool for all 
past and current photometric and spectroscopic stellar surveys. In fact, while 
it is relatively straightforward to derive some sort of information on stellar
surface temperature and chemical composition simply from colours and/or spectra
(and in many cases even detailed abundances), that is usually not the case 
when it comes to masses, radii, 
distances and, in particular, ages. Even when accurate astrometric distances 
are available to allow comparison of stars with isochrones, the derived ages 
are still highly uncertain, and statistical techniques are required to avoid 
biases \citep[e.g.,][]{PE04,JL05,S13}. Furthermore, isochrone dating is 
meaningful only for stars in the turnoff and subgiant phase 
\citep[e.g.,][]{N04,C11}, where 
stars of different ages are clearly separated in the HR diagram. This is 
in contrast, for example, to stars on the red giant branch, where isochrones 
with vastly different ages can fit equally well observational constraints such 
as effective temperatures, metallicities and surface gravities within their 
errors \citep[e.g.,][for a review]{S10}.

\section{SAGA}

The purpose of the Str\"omgren survey for Asteroseismology and Galactic 
Archaeology (SAGA) is to uniformly and homogeneously observe stars in the 
Str\"omgren $uvby$ system across the {\it Kepler} field, to derive their 
classical stellar parameters and thus provide a new benchmark for Galactic 
studies, similar to the solar neighbourhood. 
Details on survey rationale, strategy, observations and data reduction are 
provided in the first SAGA data release \citep{C14a}.

Without going into in the gory details of the Str\"omgren $uvby$ system, here 
it suffices to say that it was designed for the determination of basic stellar 
parameters with the ultimate purpose of studying Galactic stellar populations 
\citep{S63,S87}, as nicely demonstrated by the Geneva-Copenhagen Survey 
\citep[GCS,][]{N04,C11}. Indeed, SAGA builds on the legacy of the GCS, 
representing its natural extension. Similar to the latest 
revision of the GCS \citep{C11}, we combine Str\"omgren metallicities with 
broad-band photometry to obtain effective temperatures and metallicities for 
all targets via the Infrared Flux Method \citep{C06,C10,C14b}.
This facilitates the task of placing SAGA and the GCS on the same scale. 
However, there are also marked differences between the two surveys: the GCS is 
an all-sky, shallow survey limited to main-sequence and subgiant stars closer 
than $\simeq 100$~pc ($40$~pc volume limited). The {\it Kepler} targets 
observed by SAGA are primarily giants located between $\simeq1$ and 
$\simeq6$~kpc in a specific region of the Galactic disk, across the Orion arm 
and edging toward the Perseus arm.
The use of giants as probes of Galactic Archaeology is possible since it is 
relatively straightforward to derive ages for these stars once classical 
stellar parameters are coupled with asteroseismology. This was not the case 
for the GCS, where isochrone fitting was used, and thus limited to 
main-sequence and subgiant stars with known astrometric distances. On the other 
hand, stars in the GCS have kinematic information, which is not available for 
the SAGA targets. The different distance ranges sampled by the GCS and SAGA 
makes them complementary: the stellar properties measured within the solar 
neighborhood in the former survey can be dynamically stretched across 
several kpc using kinematics. In contrast, the larger distance range 
sampled by the giants in SAGA provides {\it in situ} measurements of 
various stellar properties over $\simeq5$~kpc.

Observations are being conducted with the Wide Field Camera on the $2.5$-m 
Isaac Newton Telescope (INT), which in virtue of its large field of view and 
pixel size is ideal for wide field optical imaging surveys. The purpose of SAGA 
is to obtain good photometry (i.e.~few hundredths mag) for {\it all} stars in 
the magnitude range where 
{\it Kepler} is able to measure oscillations, i.e. $10 \lesssim y \lesssim 14$. 
This requirement can be easily achieved with short exposures on a $2.5$-m 
telescope and indeed all stars for which {\it Kepler} measured oscillations are 
essentially detected in our survey (with a completeness $ \gtrsim 95$\%). 
Str\"omgren standard stars are chosen from the list of \cite{SN88}, which is 
carefully tied to the system used by \cite{O83} and underlying the GCS used 
for our previous investigation of stellar properties in the Galactic disk.

Our images are pre-processed with the Wide Field Survey Pipeline 
provided by the Cambridge Astronomical Survey Unit \citep{IL01}. The 
operations applied to the images consist of debiasing, trimming, flat 
fielding, and 
correction for non-linearity. After this pre-processing, we have developed a 
fast and efficient custom pipeline for the source detection, 
astrometric solution and photometric calibration of standard and science 
targets as detailed in \cite{C14a}.
SAGA is magnitude complete to about $y \simeq 16$~mag, and 
stars are still detected at fainter magnitudes ($y \simeq 18$), although 
with increasingly larger photometric errors and incompleteness. As part of 
SAGA we are also obtaining photometry on the brightest targets using the 
four-channel photometer at San Pedro M\'artir Observatory.

With SAGA it is thus straightforward to build a magnitude complete and unbiased 
photometric catalog down to $y \simeq 16$~mag, against which we can benchmark 
the 
sample of stars with measured {\it Kepler} oscillations. This makes our 
photometric survey unique in terms of recovering the {\it Kepler} selection 
function of seismic targets \citep{C15}.  
In fact, the selection criteria of the {\it Kepler} mission were designed to 
optimize the scientific yield of the mission with regard to the detection of 
Earth-size planets in the habitable zone of stars \citep{B10}. Thus, while the 
selection function is known for exoplanetary studies, this is not the case 
when it comes seismic targets, entries in the 
seismic sample of giants being based on a number of heterogeneous criteria 
\citep[e.g.,][]{H10}. Our approach thus complements other ground based follow
up studies of asteroseismic targets \citep[see e.g.][for the rationale behind 
photometric parameters and a brief discussion of pros and cons between 
photometric and spectroscopic surveys]{CV,C14c}. In future observing runs we 
plan to extend some of our pointings also to {\it K2} fields \citep{K2}.

\begin{figure}
\resizebox{0.99\columnwidth}{!}{%
\includegraphics{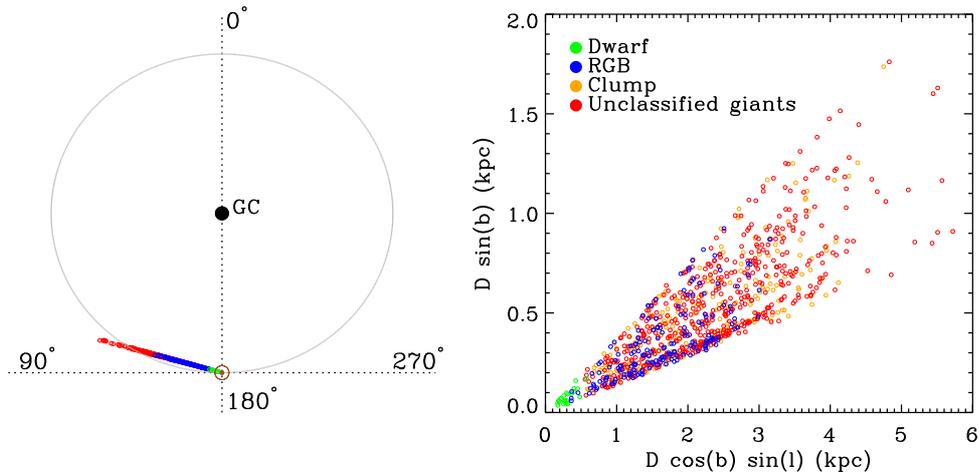}}
\caption{Location of SAGA targets in the Galaxy. Stars with different seismic 
classification have different colours, as labelled. {\it Left panel:} 
face-on view in Galactic coordinates, where the distance of each seismic 
target from the Sun ($D$) is projected along the line of sight $D\,\cos(b)$ 
having direction $l\simeq 74^{\circ}$ and Galactic latitude $b$. The distance 
between the Galactic Centre (GC) and the location of the Sun ($\odot$) is 
marked by the solar circle (in gray). Galactic longitudes ($l$) at four 
different angles are 
indicated. {\it Right panel:} same stars as function of Galactic height 
$\textrm{Z} = D\,\sin(b)$ and projected across the $l=90^{\circ}$ 
direction.}
\label{fig:1}
\end{figure}

\section{Stellar parameters and Galactic structure}

The first SAGA data release combines asteroseismic and photometric stellar 
parameters for 989 {\it Kepler} targets, most of which are red giants. The 
location of the targets in the Galaxy is shown in Figure \ref{fig:1}. Within 
SAGA a novel approach is developed to derive classical and asteroseismic 
stellar parameters in a fully self-consistent way: for each target, its 
photometric effective temperature and metallicity, together with the mass, 
radius, surface gravity, density and distance is computed. For a large 
fraction of objects, evolutionary phase classification tells whether a stars 
is a dwarf, is evolving along red giant branch (RGB) or is already in the clump 
phase \citep[e.g.][]{Stello13}. The latter distinction is particularly 
important to derive reliable stellar ages, in particular to select only 
{\it bona-fide} lower RGB stars, where the effect of mass loss is negligible 
\citep[e.g.][]{VSA14}.

We provide a careful assessment of random and systematic uncertainties on our 
parameters. Total uncertainties are of order $82$~K in effective temperature, 
$0.17$~dex in metallicity, $0.006$~dex in surface gravity, $1.5$\% in stellar 
density, $2.4$\% in radius, $3.3$\% in distance and $6.0$\% in mass. Age 
uncertainties vary depending an the availability of seismic classification or 
not, but are usually below $30$\% \citep{C14a,C15}.
Confidence in the achieved precision is corroborated by the detection of the 
first and secondary clumps in a population of field stars and by the 
negligible scatter in the seismic distances and ages among NGC\,6819 member 
stars (one of the four open clusters located in the {\it Kepler} field).

With the stellar parameters derived so far, we are thus in the position of 
using {\it Kepler} targets, and asteroseismology, to investigate some of the 
most important constraints on Galactic models, such as the age-metallicity 
relation, and the vertical structure of the Galactic disk via age, mass and 
metallicity gradients \citep{C15,S15}. Furthermore, calibrating photometric 
metallicities and age-dating techniques for the entire photometric sample, the 
continuation of our Str\"omgren survey promises a leading role for Galactic 
studies.

\end{document}